# Chemical Tuning of Positive and Negative Magnetoresistances, and Superconductivity in 1222-type Ruthenocuprates

Abbie C. Mclaughlin,*,† Laura Begg,† Catriona Harrow,† Simon A. J. Kimber,§ Falak Sher§ and J. Paul Attfield*,§

*The Chemistry Department, University of Aberdeen, Meston Walk, Aberdeen, AB24 3UE,*
*Centre for Science at Extreme Conditions and School of Chemistry, University of Edinburgh, King's Buildings, Mayfield Road, Edinburgh, EH9 3JZ.*

High critical-temperature superconductivity and large ('colossal') magnetoresistances are two important electronic conducting phenomena found in transition metal oxides. High-$T_c$ materials have applications such as superconducting magnets for MRI and NMR, and magnetoresistive materials may find use in magnetic sensors and spintronic devices. Here we report chemical doping studies of $RuSr_2(R_{2-x}Ce_x)Cu_2O_{10-\delta}$ ruthenocuprates which show that a single oxide system can be tuned between superconductivity at high hole dopings and varied magnetoresistive properties at low doping levels. A robust variation of negative magnetoresistance with hole concentration is found in the $RuSr_2R_{1.8-x}Y_{0.2}Ce_xCu_2O_{10-\delta}$ series, while $RuSr_2R_{1.1}Ce_{0.9}Cu_2O_{10-\delta}$ materials show an unprecedented crossover from negative to positive magnetoresistance with rare earth ($R$) ion radius.

Although the mechanism for superconductivity in layered cuprates remains controversial,[1] the chemical tuning of their properties is well-established. Oxidation of the $CuO_2$ planes suppresses antiferromagnetic order of $Cu^{2+}$ $S= ½$ spins, and induces superconductivity in the doping range p= 0.06-0.25 (the equivalent Cu oxidation states are 2+p). Ruthenocuprates contain distinct $RuO_2$ and $CuO_2$ planes, and display coexisting ferromagnetism and superconductivity in both 1212-type ($RuSr_2RCu_2O_8$)[2,3] and 1222-type ($RuSr_2(R,Ce)_2Cu_2O_{10-\delta}$)[4,5] structures, where $R$= Sm, Eu, or Gd. Large negative magnetoresistances (change of electrical resistivity $\rho$ in an applied magnetic field H, defined as MR= ($\rho$(H) - $\rho$(0))/$\rho$(0)) have recently been observed in non-superconducting $R$ = (Nd,Y) 1222 materials[6], up to MR = -34% for $RuSr_2NdY_{0.1}Ce_{0.9}Cu_2O_{10-\delta}$ at 4 K and 7 T. The results presented here reveal an exquisite chemical tuning of the properties of 1222 ruthenocuprates between superconducting and magnetoresistive properties, including a previously unreported, large positive magnetoresistance state.

1222-type ruthenocuprates contain metal oxide layers in the repeat sequence; ..$RuO_2$.$SrO$.$CuO_2$.$(R,Ce)$.$O_{2-\delta}$.$(R,Ce)$.$CuO_2$.$SrO$.. (Table of Contents figure shows the crystal structure). The chemistry of the $(R,Ce)_2O_{2-\delta}$ slab between the two $CuO_2$ planes controls the electronic properties. Two series of polycrystalline ceramic 1222 samples, $RuSr_2Nd_{1.8-x}Y_{0.2}Ce_xCu_2O_{10-\delta}$ (0.70 < x < 0.95) and $RuSr_2R_{1.1}Ce_{0.9}Cu_2O_{10-\delta}$ ($R_{1.1}$= $Nd_{1.0}Y_{0.1}$, $Nd_{0.9}Y_{0.2}$, $Sm_{1.1}$, $Sm_{0.9}Y_{0.2}$, $Eu_{0.9}Y_{0.2}$, $Gd_{0.9}Y_{0.2}$; in order of decreasing radius), were prepared to determine the respective effects of variable doping and $R$ cation size. Pelleted stoichiometric mixtures of $R_2O_3$, $RuO_2$, CuO, $CeO_2$ and $SrCO_3$ powders were repeatedly sintered at 1025 ºC and furnace cooled in air. Tetragonal $I4/mmm$ 1222 phases ($a \approx$ 3.85, $c \approx$ 28.55 Å) were formed in all cases.

It is difficult to prepare single phase Nd-based 1222 materials as these only form over a ~10 ºC synthesis window around 1025 ºC, and a small degree of Y substitution was needed to produce samples having a high degree (>95%) of phase purity. Portions of high (low) x $RuSr_2Nd_{1.8-x}Y_{0.2}Ce_xCu_2O_{10-\delta}$ samples were annealed under flowing $N_2$ ($O_2$) to increase (decrease) the oxygen deficiency $\delta$ and thereby extend the available hole-doping range [7]. Precise oxygen stoichiometries were determined by thermogravimetric reduction under 5% $H_2$ in $N_2$. Ru is known to be in the +5 state in 1222 materials from previous XANES and doping studies [8], giving p = (1 − x − 2$\delta$)/2. Ten $RuSr_2Nd_{1.8-x}Y_{0.2}Ce_xCu_2O_{10-\delta}$ samples were synthesised, with 0.70 < x < 0.95 and 0.004 < $\delta$ < 0.095, giving a doping range 0.010 < p < 0.059 that spans the presuperconducting region for cuprates. Further sample details are given in Supporting Information.

All of the $RuSr_2Nd_{1.8-x}Y_{0.2}Ce_xCu_2O_{10-\delta}$ materials are semiconducting with bandgaps of 30-170 meV and the Cu spins order antiferromagnetically at 25-115 K. The magnetoresistance properties are typified by the data for the x = 0.9 sample in Fig. 1(a) and (b). Negative MR is observed below the Ru spin ordering transition ~160 K, diverging to large values below the Cu spin ordering temperature of 60 K, and the 4 K MR varies near-linearly with magnetic field. These properties are characteristic of charge transport by magnetopolarons – small ferromagnetic regions surrounding each Cu-hole within a matrix of antiferromagnetically ordered $Cu^{2+}$ spins.[9] An applied magnetic field cants the Ru spins into a ferromagnetic arrangement, which induces partial ferromagnetism in the $CuO_2$ planes thereby increasing the mobility of the magnetopolarons, giving the observed, negative MR's [6]. Magnetopolaron hopping is a thermally activated process, leading to a characteristic exponential rise in –MR below the Cu spin transition,[10] as seen in Fig. 1(a).

The magnetoresistances in the $RuSr_2Nd_{1.8-x}Y_{0.2}Ce_xCu_2O_{10-\delta}$ series (at 4 K and 7T) shows a striking correlation with the doping level p, as shown in Fig. 2. No such MR trend has been reported previously in any cuprate series. -MR initially rises with p, reflecting the increasing number of hole carriers in the $CuO_2$ planes, but falls sharply above p = 0.04. The known onset of superconductivity in ruthenocuprates (and cuprates in general) at p> 0.06, suggests that this collapse of –MR corresponds to the onset of hole-pairing, as the singlet Cu hole pairs that carry the supercurrent do not contribute to the magnetotransport. Hence, our magnetoresistance data show that superconducting pair formation starts above a distinct threshold concentration of 4% doping, although the coherent superconducting state is observed only for >6% doping.

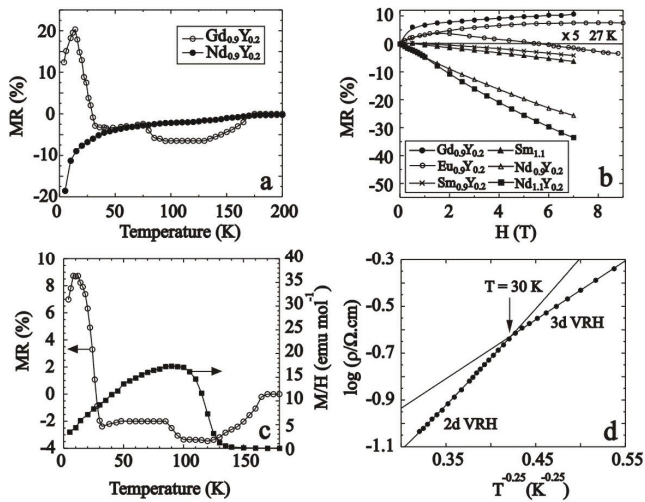

**Figure 1.** Magnetoresistance (MR) data for $RuSr_2R_{1.1}Ce_{0.9}Cu_2O_{10-\delta}$ materials. (a) Temperature variations of MR in a 5 T field for two representative samples. (b) MR variations with magnetic field at 4 K for all samples; also for $R_{1.1} = Eu_{0.9}Y_{0.2}$ at 27 K, scaled x 5. (c) Temperature variations of magnetoresistance (H= 5 T) and magnetic susceptibility (H= 0.01 T) for $R_{1.1} = Eu_{0.9}Y_{0.2}$. (d) Plot of log(resistivity) against $T^{-1/4}$ for $R_{1.1} = Eu_{0.9}Y_{0.2}$ between 12 and 95 K, showing the change from 3-dimensional to 2-dimensional variable range hopping conductivity at 30 K.

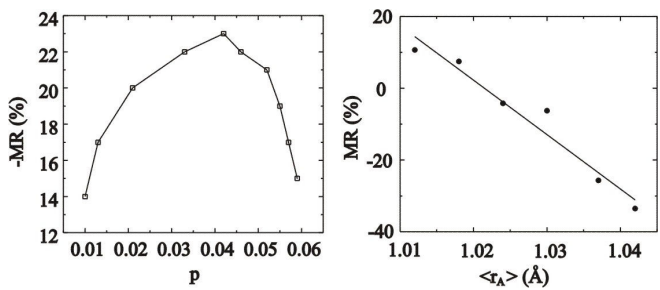

**Figure 2.** (Left) Variation of the $-MR_{7T}(5\ K)$ magnetoresistance with hole doping p in the $RuSr_2Nd_{1.8-x}Y_{0.2}Ce_xCu_2O_{10-\delta}$ series. (Right) Variation of $MR_{7T}(4\ K)$ with $<r_A>$, the mean A (=$R_{1.1}Ce_{0.9}$) site cation radius.

To discover how $R^{3+}$ cation size influences magnetoresistance at a constant R/Ce ratio, a series of $RuSr_2R_{1.1}Ce_{0.9}Cu_2O_{10-\delta}$ samples was prepared. A remarkable change is observed on going from the above R= (Nd,Y) materials to the small radius R= (Gd,Y) and (Eu,Y) analogs, which show a complex thermal evolution of MR (in Figs. 1(a) and (c), respectively). Cooling below the onset of negative MR, a small discontinuity to less negative values is seen at the Cu spin ordering transition at 80 K. This is identified from the susceptibility maximum in Fig. 1(c) [6], and suggests that the Cu spin structure is different to that in the R= (Nd,Y) materials. A sharp transition at 30 K leads to a large positive MR peak of 22% at 10 K for R= (Gd,Y). The low temperature, positive MR rises rapidly with field, and higher temperature (27 K) measurements show that MR goes through a broad maximum and then decreases to negative values (Fig. 1(b)). The 4 K MR's in the $RuSr_2R_{1.1}Ce_{0.9}Cu_2O_{10-\delta}$ series shows an approximately linear crossover from large positive to negative values with increasing R cation radius[11]. There is no corresponding correlation between MR and the $R^{3+}$ paramagnetic moment, confirming that the size effect is dominant. Hence, the 1222 ruthenocuprates show a novel positive to negative MR transition that can be induced by increasing temperature, magnetic field strength, or cation radius. A thermal crossover from negative to positive MR has been reported in $Zn_{1-x}Cu_xCr_2Se_4$ selenides [12], but the strong size control of the sign of MR found here is unprecedented.

No magnetic anomaly is observed at the 30 K MR discontinuity in $RuSr_2Eu_{0.9}Y_{0.2}Ce_{0.9}Cu_2O_{10-\delta}$ (Fig. 1(c)), however, a subtle electronic transition is evidenced from a change in the resistivity variation (Fig. 1(d)) from $\rho \sim \exp(A.T^{-1/3})$ above 30 K to $\exp(A.T^{-1/4})$ below. These correspond to variable-range hopping of the carriers in two and three dimensions, respectively. Hence, the negative MR above 30 K is associated with two-dimensional transport in the $CuO_2$ planes, whereas below 30 K, additional conduction between planes has a different, positive MR mechanism that dominates the bulk MR of ceramic samples. The correlation between low temperature MR and mean cation radius in Fig. 2 thus results from the decrease in the width of the insulating $(R,Ce)_2O_{2-\delta}$ slab between $CuO_2$ planes as cation size decreases, leading to more three-dimensional conducting behaviour in the small cation materials.

In summary, the combination of electronically active $RuO_2$ and $CuO_2$ planes and a sophisticated chemical tuning from the $(R,Ce)_2O_{2-\delta}$ slabs leads to an exquisite variation of low temperature conducting properties in $RuSr_2(R_{2-x}Ce_x)Cu_2O_{10-\delta}$ materials. Low-doped, large $R^{3+}$ materials such as R = (Nd,Y) show large negative MR's characteristic of two-dimensional magnetopolaron hopping, suppressed at higher dopings by superconducting pair formation. Analogs with small R = Eu or Gd have positive MR's associated with three-dimensional hopping, crossing over to negative MR's at high temperatures or fields. Superconducting materials with $T_c$'s up to 50 K can also be prepared using high R = Sm, Eu, or Gd contents [4, 5, 13, 14]. Further studies of these materials may give new insights into superconducting and magnetoresistive electron transport in oxides, and could lead to improvements in devices such as spin-polarised quasiparticle junctions.[15]

**Acknowledgment** The Royal Society of Edinburgh for a SEELLD research fellowship (ACM), the Ministry of Science and Technology, Government of Pakistan for a studentship (FS), and the UK EPSRC for beam time provision and financial support.

**Supporting information available:** Table of $RuSr_2Nd_{1.8-x}Y_{0.2}Ce_xCu_2O_{10-\delta}$ compositions.

**References**

[†] University of Aberdeen.
[§] University of Edinburgh
(1) Reznik, D.; Pintschovius, L.; Ito, M.; Iikubo, S.; Sato, M.; Goka, H.; Fujita, M.; Yamada, K.; Gu, G. D.; Tranquada, J. M. *Nature* **440**, 1170-1173 (2006).
(2) Mclaughlin, A. C.; Janowitz, V.; McAllister, J. A.; Attfield, J. P. *Chem. Commun.* **2000**, 1331-1332.
(3) Mclaughlin, A. C.; Janowitz, V.; McAllister, J. A.; Attfield, J. P. *J. Mater. Chem.* **2001**, 11, 173-178.
(4) Knee, C. S.; Rainford, B. D.; Weller, M. T. *J. Mater. Chem,* **2000**, 10, 2445-2447.
(5) Mclaughlin, A. C.; Attfield, J. P.; Asaf, U.; Felner, I. *Phys. Rev. B* **2003**, 68, 014503.
(6) Mclaughlin, A. C.; Sher, F.; Attfield, J. P. *Nature* (London) **2005**, 436, 829-832; **2005**, 437, 1057-1057.
(7) Samples were annealed under flowing $N_2$ at 600°C or under flowing $O_2$ at 800 °C and then furnace cooled.
(8) Williams, G. V. M.; Jang, L.-Y.; Liu, R. S. *Phys. Rev. B* **2002**, 65, 064508.
(9) Nagaev, E. L. *JETP Lett.* **1967**, 6, 18-.
(10) Majumdar, P.; Littlewood, P. *Phys. Rev. Lett.* **1998**, 81, 1314-1317.
(11) Shannon, R. D. *Acta Cryst. A.* **1976,** 32, 751-767.
(12) Parker, D. R.; Green, M. A.; Bramwell, S. T.; Wills, A. S.; Gardner, J. S.; Neumann, D. A. *J. Am. Chem. Soc.* **2004**, 126, 2710-2711.
(13) Ono, A. *Jpn. J. Appl. Phys.* **1995**, 34, L1121.
(14) Felner, I.; Asaf, U.; Levi, Y.; Millo, O. *Phys. Rev. B* **1997**, 55, R3374-R3377.

(15) Dong, Z. W.; Pai, S. P.; Ramesh, R.; Venkatesan, T.; Johnson, M.; Chen, Z. Y.; Cavanaugh, A.; Zhao, Y. G.; Jiang, X. L.; Sharma, R. P.; Ogale, S.; Greene, R. L. *J. Appl. Phys.* **1998**, 83, 6780-6782.

TOC:

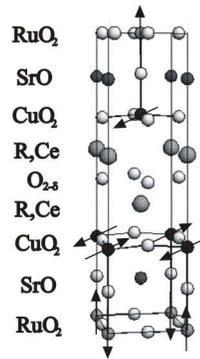

A remarkable variety of conducting states has been found in RuSr$_2$($R_{2-x}$Ce$_x$)Cu$_2$O$_{10-\delta}$ ruthenocuprates by tuning the properties of the magnetic CuO$_2$ and RuO$_2$ layers through small changes in the chemistry of the ($R$,Ce)$_2$O$_{2-\delta}$ slab. Both the $R^{3+}$ cation size and the charge transfer determined by the R/Ce ratio and the oxygen deficiency $\delta$ are important controlling parameters that tune ground state properties from positive magnetoresistive to negative magnetoresistive to superconducting.

ABSTRACT FOR WEB PUBLICATION (Word Style "BD_Abstract"). Authors are required to submit a concise, self-contained, one-paragraph abstract for Web publication.